\documentclass[letterpaper]{JHEP3}

\usepackage{epsfig, epstopdf}

\def\be{\begin{equation}}
\def\ee{\end{equation}}
\def\bea{\begin{eqnarray}}
\def\eea{\end{eqnarray}}
\def\nn{\nonumber}

\def\exd{{\rm d}}

\def\endignore{}
\def\ignore #1\endignore{} 
\def\ol#1{\overline{#1}}

\def\bd{\begin{displaymath}}
\def\ed{\end{diplaymath}}
\def\hR{{\hat{R}}}
\def\bR{{\breve R}} 
\def\x{{\hat \xi}}     

\def\cF{{\cal F}}
\def\ps{{ \tilde \psi}}

\def\ba{\begin{eqnarray}}
\def\ea{\end{eqnarray}}
\def\be{\begin{equation}}
\def\ee{\end{equation}}

\def\exd{{\rm d}}

\title{Dynamics of D3-D7 Brane Inflation in Throats
}

\author{
Fang Chen$^{1} $\footnote{fangchen@hep.physics.mcgill.ca} ,   Hassan Firouzjahi$^{2}$\footnote{firouz@ipm.ir}
\\
$^1$ Physics Department, McGill University,  Montreal, H3A 2T8, Canada. \\
$^{2}$ School of Physics, Institute for Research in Fundamental Sciences (IPM),\\
~~~~~~~~~~~~~~~ P.O.Box 19395-5531,
Tehran, Iran
}

\date{}

\abstract {
Dynamics of D3-branes in models of warped D3-D7 inflationary set up is studied
where perturbative $\alpha'$ correction to the K\"ahler potential and
the nonperturbative corrections to the superpotential are included.
It is shown that a dS minimum can be obtained without introducing
anti-branes. Some specific configurations of D7-branes embedding were
studied. After stabilizing the angular directions,  it is shown that
the resulting D3-D7 potential of the radial position of the D3-brane
is too steep to allow slow-roll inflation. Depending on D7-branes embedding and the
stabilized angular directions, the mobile D3-brane can move either towards the tip of
the throat or towards the D7-branes.
}

\begin{document}

\section{Introduction} Obtaining metastable de Sitter vacua and
inflation within  type IIB string theory has been a major focal point
of string cosmology in recent years.  Compactifications making use of
warped throats \cite{Klebanov:2000hb,GKP} have provided a self-consistent starting
point, but the simplest versions lead to AdS rather than Minkowski or
dS minima.  The introduction of anti-D3 branes into the throat can 
uplift the AdS minimum \cite{KKLT}, and can allow for brane-antibrane
inflation \cite{KKLMMT}.  

However the antibranes explicitly break supersymmetry, while the
low-energy description of the K\"ahler moduli and inflaton field is
in the language of supergravity.  This tension has motivated the
search for other uplifting mechanisms which are not explicitly
SUSY-breaking, notably magnetic fluxes on the D7-branes, leading to
D-term uplifting \cite{BKQ}.  Other examples include ref.\
\cite{BCDF}, in which it is pointed out that the back reaction
of D7-branes leads to a correction of the dilaton  background that
can give uplifting in the presence of D3 branes.

The present work is motivated by the possible interplay between the
mechanism of uplifting and the naturalness of brane-antibrane
inflation \cite{dvali-tye, collection}
in the warped throat( for an extensive review on brane inflation see \cite{HenryTye:2006uv}  and references therein). These two issues are related through
the infamous $\eta$ problem \cite{KKLMMT}: because the D3-brane
modulus (which is also the inflaton) $\phi$ enters into the K\"ahler
potential $K$ through the combination $R = 2\sigma (1 - c\phi^2/M_p^2)$ 
(where $\sigma$ is the K\"ahler modulus and $c\sim 1$) and all terms
in the inflaton potential go like $1/R^p$ with $p$ being a positive
power, the inflaton mass is naturally of order $V/M_p^2 = H^2$, and
so the inflationary $\eta$ parameter is $O(1)$.  The inflaton mass
can be made small only by finely-tuned competing contributions,
notably the $\phi$ dependent corrections to the nonperturbative
superpotential \cite{BCDF, Firouzjahi:2003zy, Shandera:2004zy, McAllister:2005mq,
Baumann1,Baumann2}.  

As in \cite{GKP}, we assume that all 
complex structures are stabilized by turning a variety of background fluxes. The resulting classical superpotential is independent of the K\"ahler modulus leading to a
no-scale potential  where the K\"ahler modulus remains completely undetermined. 
The no-scale property of the potential can be broken either via non-perturbative corrections to
the superpotential as employed in \cite{KKLT}, or via perturbative $\alpha'$ correction into the K\"ahler potential as demonstrated in \cite{BBHL}.  In  \cite{BBHL} it is shown that the leading
corrections are in the form of $K =  -2\ln(R^{3/2} + \xi)$ where $\xi$ represents
the $\alpha'^3$ correction to the K\"ahler potential. Uplifting using this $\alpha'$ correction into the K\"ahler potential was discussed previously in  \cite{BB, Cremades, Misra:2007yu}. We also find that in our D3-D7 
setup uplifting can be achieved from the interplay of leading  perturbative $\alpha'$ correction and non-perturbative superpotential corrections, without any
antibrane.  Since the higher order $\alpha'$ correction into the K\"ahler potential are not known, we work in a large volume limit where only the leading perturbative $\alpha'$ correction into the K\"ahler potential are trusted. Also we work in the limit where  the constant term in the superpotential $W_0$ to be  much larger than in the KKLT \cite{KKLT} proposal, which
is a virtue since the KKLT value is unnaturally small.

Having obtained a dS vacuum, we add the D3-brane into the setup. As in \cite{Baumann1}, we 
take into account the D3-brane corrections into the non-perturbative superpotential. This results in a potential for the  mobile D3 brane in the throat.  To study
its suitability for inflation, we must minimize the potential with
respect to the angular directions in the throat for a given  radial
position $r$, resulting in the effective single-field inflaton
potential (since $\phi\propto r$).  We perform this minimization for
two characteristic embeddings of the D7 branes in the throat, the
Ouyang and Kuperstein embeddings \cite{Ouyang:2003df, Kuperstein:2004hy} .  In neither case can  $V(\phi)$ be
made flat enough to give inflation.  The problem can be traced to
the requirement of large $W_0$.  The nonperturbative superpotential
corrections, which were tuned to cancel the contributions to $\eta$
coming from $W_0$ in \cite{Baumann2}, cannot be made sufficiently
large to enable this cancellation when $|W_0| \sim 1$.  

We begin in section \ref{sugra4d} with the 4D supergravity effective action
describing the K\"ahler modulus and D3 brane modulus, including the
effect of $\alpha'$ corrections in the K\"ahler potential.  Here we
give a detailed realization of the uplifting due to such corrections.
In section \ref{conifold} we review the description of the geometry of the
conifold to introduce the angular degrees of freedom of the D3-brane
in the throat.  These angles are minimized at different values,
depending on the embedding of the D7 branes and the radial position of
the D3 in the throat.  In section \ref{embeddings} we determine these
angular minima for two representative D7 embeddings which have been
discussed in previous literature.  For each case we show that the
potential for the radial modulus of the D3-brane cannot be tuned to be
flat enough for inflation.  In the final section \ref{conclusions}, we argue that this is
a generic feature of large volume compactifications.


\section{The low energy 4D supergravity}
\label{sugra4d}

We start with the 4D effective supergravity for the dynamics of
two light fields: $T $, describing the volume
modulus and $z^i$ denoting the position in the extra
dimensions of the D3 brane. We suppose the dynamics of these fields to
be governed by gaugino condensation on the D7 brane, in the usual
KKLT fashion, including the back-reaction of the D3 brane as
computed by Baumann et.~al.~\cite{Baumann1}.

\subsection{General definitions}

The K\"ahler potential, $K$, and superpotential, $W$, defining the low-energy 4D supergravity are given by \cite{BBHL}
\bea
\label{KW}
    K &=& -2 \ln( R^{3/2}+\xi) 
    \nn\\
    W &=& W_0 + A(z) \,  \exp[-b \, T] \, .
\eea
where $ T= \sigma + i \tau $  is the complex K\"ahler modulus, $\tau$ is the axion-dilaton field and $\sigma$ is the volume modulus. More specifically, 
$\sigma^{1/4}$ measures the size of the CY compactification
\ba
\label{sigma}
\sigma \sim \frac{R_{CY}^{4}}{\alpha'^{2}} \, .
\ea
Here $R_{CY}$ represents the average radius of the CY compactification. To trust low energy
supergravity formalism, we require $\sigma \gg1$. We take $\sigma \sim  {\cal O}(100)$ in our discussions below.
Also 
\be
    R = T+\ol{T} - \gamma k(z,\ol{z})
    = 2\sigma - \gamma k(z,\ol{z}) \,. 
\ee
Here $k(z,\ol{z})$ denotes the K\"ahler potential for the
CY geometry with the K\"ahler metric $k_{i \bar j}=\partial_{i} \partial_{\bar j} k(z,\bar z)$
and $z^{i}, i=1,2,3$ represent the position of the mobile
D3-brane inside CY compactification. Also $A(z) = A_0 [F(z)]^{1/n}$ where the
holomorphic condition $F(z) = 0$ defines the position of the stack of $n$ D7
branes in this geometry. The exponential term in $W$ represents the non-perturbative correction to the superpotential coming from the gaugino condensation on coincident D7-branes and $b=2\pi/n$.
Finally, $\gamma$ is related to the tension of the D3-brane
and is given by 
\ba
\label{gamma}
\gamma = \frac{\sigma}{3} \frac{T_{3}}{M_{P}^{2}} \, 
\ea
where  $T_{3} \sim m_{s}^{4}$ is the tension of the D3-brane.

The F-term potential is computed from $K$ and $W$ using the
standard formula
\ba
\label{VF}
 V_F &=& e^K \Bigl[ K^{\ol{a}b} \ol{D_a W} D_b W - 3 |W|^2 \Bigr]  \nonumber\\
   &=& e^{K} \left[ (K^{\bar a b}K_{\bar a} K_{b} -3) |W|^{2} + K^{\bar a b}  \ol{W}_{a}  W_{b} + 
 K^{\bar a b} W  \ol{W}_{a}  K_{b}  + K^{\bar a b} \ol{W}  {W_{b} } { K}_{\bar a}  \right]  \, ,
\ea
where $D_a W = W_a + K_a W$ and $K^{\ol{a}b}$ denotes the inverse, $K^{\ol{a}b} K_{b\ol{c}}
= {\delta^{\ol{a}}}_{\ol{c}}$, of the K\"ahler metric on moduli
space, $K_{b\ol{c}} = \partial_b \partial_{\ol{c}} K$.


\subsection{Evaluation of the potential}

We now explicitly evaluate the derivatives appearing in the above
definitions, using the given functions $K$, $W$. The
superpotential derivatives become
\be
    W_T = -b A  \exp(-bT) \,, \quad
       W_i = A_i  \exp(-bT)\,, 
\ee
where the superscript $i$ represents the partial derivative with respect to $z^{i}$ the
position of the mobile D3-brane.
It is useful to define  $ \hat \xi = \xi  \, R^{-3/2} $ and $\hat R = R (1+\hat \xi)$
where the K\"ahler potential is now given by 
\ba
\label{k1}
K=  -2 \ln (R^{1/2}\hat R)  \, .
\ea
In this notation, the quantity $\x$ measures the strength of $\alpha'$ correction to the K\"ahler potential. The consistency of our perturbation requires that $\x \ll 1$.

The K\"ahler derivatives are given by
\be
    K_T = - \frac{3}{ \hat R  }\,, \quad
    K_i = \frac{3\gamma k_i}{ \hat R} \,,
\ee
where here and below $k_{i} \equiv \partial_{i} k$ and $k_{\bar i} = \partial_{\bar i} k$. 

The K\"ahler derivatives of $W$ evaluate to
\bea
    D_T W &=& -  \frac{3 W_{0} }{ \hat R }  - A  \exp(-bT) \left( b +
    \frac{3}{    \hat R }  \right) \nn\\
      D_i W &=& \frac{3\gamma W_0 k_i}{\hat R  } + A  \exp(-bT)
    \left( \frac{A_i}{A} + \frac{3\gamma k_i}{ \hat R } \right) \,. \nn
\eea
The K\"ahler metric is given by
\be
    K_{a\ol{b}} =  \frac{3  }{ \hat R   \breve R  }
    \left(%
    \begin{array}{ccc}
    1  & - \gamma\, k_{\ol\jmath} \\
    -\gamma\,  k_i  ~~ \, \, &       \gamma \, \breve R       k_{i\ol\jmath} + \gamma^{2} \, k_i k_{\ol\jmath}  \\
    \end{array}%
    \right) \,,
\ee
where 
\be \label{cR}
 \breve R \equiv \frac{2(1+\hat \xi ) }{  (2-\hat \xi ) } R \, .
\ee
After some algebra, the inverse metric is
\be
    K^{\ol{a} b} = \frac{ \hat R   }{3} \left(%
    \begin{array}{ccc}
    \breve R +\gamma  k^{\bar i} k_{\bar i} & k^i \\
    k^{\ol\jmath} &  \, \gamma^{-1}\, k^{\ol\jmath i}  \\
    \end{array}%
    \right) \,,
\ee
where $k^{\bar i j}$ is the inverse of $k_{\bar i j}$ with $k^{\bar i j} k_{j \bar l} = \delta^{\bar i}_{\bar l}$ and $k^{ i} = k_{  \bar j} k^{\bar j i} $.

Calculating the F-term potential from (\ref{VF}), one obtains
\ba
\label{totalVF}
V_{F}= \frac{1}{ 3 R \hat R  } \left[   |A|^{2} e^{- 2 b \sigma} 
 \left( (\bR +\gamma  k^{ i} k_{ i}  ) b^{2} + 6 b \frac{\bR}{\hR} \right)
  + 3 b W_{0}  \frac{\bR}{\hR} (A e^{-b T} + c.c.)  \right. \nonumber\\
\left.  -  b\, e^{-2 b \sigma} (A \, k^{\bar i} \bar A_{i} + c.c.)
+ \gamma^{-1} e^{-2 b \sigma} k^{\bar i j} \bar A_{i} A_{j} + \frac{9}{\hR^{2}} (\bR - \hR) |W|^{2}
\right] \, .
\ea
As in \cite{BBHL, BB} the no-scale
property of the four-dimensional supergravity is broken in the
presence of $\xi$. This manifests itself in the last term containing
$|W|^{2}$.

We note that there are two terms in potential (\ref{totalVF}) containing the axion field $\tau$ in the form of  $W_{0}( A e^{-i b \tau} + c.c.)$. Minimizing the potential with respect to $\tau$ and noting 
that $W_{0}<0$, the stable value of axion is given by
\ba
e^{- i b \tau} =\left( \frac{\bar A}{A} \right)^{1/2} \, .
\ea
As mentioned in \cite{Baumann2}, plugging this value of the axion field into the 
potential (\ref{totalVF}), is equivalent to 
replacing  $( A e^{-i b \tau} + c.c.)$ by $2 |A|$, which we consider is the case from now on.


\subsection{Uplifting}

In this section we study the uplifting and the stabilization in the absence of the mobile D3-brane. For uplifting in this mechanism to be achieved, we need to work in the limit where $|W_{0}| \sim 1$ which is natural in flux compactification.
This is opposite of the KKLT limit  where $|W_{0}|  \ll 1$ by tuning.

An analytical understanding of the background is very helpful
when we include the D3-brane in next section in search for getting inflation.
In the absence of D3-brane, $W_{i}= k_{i}= 0$, $R = 2 \, \sigma$ and $A = A_0$.  
Eliminating the axion field as mentioned before,  the F-term 
potential for $\sigma$ obtained from (\ref{totalVF}), in the convention $A_{0}=1$, is
\ba
\label{VF1}
V_{F} = \frac{1}{ 24 \sigma^{3} (1+\x)^{2} (2- \x)     } 
\left[ \,   9 W_{0}^{2} \x + 6 A_{0} W_{0} e^{-b \sigma} \left( 4 b \sigma ( 1+ \x)  + 3 \x  \right)    \right.\nonumber\\
\left. + A_{0}^{2} e^{-2 b \sigma} \left(\,   8 b \sigma( 3+ b \sigma) + ( 3+ 4 b \sigma)^{2} \x 
+ 8 b^{2} \sigma^{2} \x^{2}  \,  \right)\, 
\right] \, .
\ea
As explained before, the breakdown of the no-scale symmetry manifests itself in the
term containing $W_{0}^{2}$, corresponding to the zeroth order contribution of the superpotential to
the potential.

In order to get a better understanding of the uplifting and the K\"ahler modulus stabilization mechanism, here we expand the potential up to quadratic order in terms of  $\x=\xi/(2 \sigma)^{3/2}$
\ba
\label{VFxi2}
V_{F} \simeq \frac{b W_{0}}{ 2 \sigma^{2}}   e^{- b \sigma} 
+ \frac{3 W_{0}^{2} \xi }{  2^{11/2}  \,  \sigma^{9/2}  } 
- \frac{9 W_{0}^{2} \xi^{2} }{  256\,  \sigma^{6} }  \, .
\ea
We assume that  that $W_{0}$ is large enough, possibly ${\cal O} (10)$, such that terms containing $e^{-2 b \sigma}$ are negligible.
The  details of  the conditions for uplifting and stabilization of the K\"ahler modulus
is relegated to the { \bf Appendix \ref{appA}}. In the limit of small $\x$, we find the following approximation
relations
\ba
\label{upapp}
b \sigma \simeq 3  \quad , \quad | W_{0}| \xi b^{3/2} \simeq 6 \, .
\ea
Using the relation $b=2 \pi/n$, we find that
$\sigma \simeq n/2$ and $W_{0} \sim \x^{-1}$.
For uplifting to work in this mechanism we need to start withe large enough $|W_{0}|$ as mentioned before. Also to obtain a large enough compactification, we should start with a large enough number of wrapped D7-branes.

\begin{figure}[t] 
\vspace{-0.8cm}
   \centering
   \includegraphics[width=3.5in]{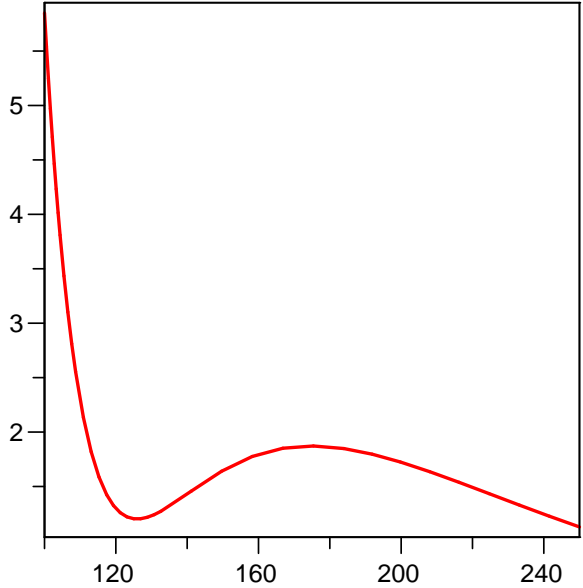} 
   \caption{In this figure, $V_{F}$ versus $\sigma $ is plotted for $A_{0}=1, W_{0}=-8, n=314$ with $\xi=300$.  The vertical line is in units of $10^{-8}$.}
\label{VFxi}
\end{figure}

Here we provide some examples where potential (\ref{VF1}) can be stabilized to a dS vacuum.
For the first example, suppose $A_{0}=1$, $W_{0}=-8$ and $n=157$, corresponding to $b= 0.04$.
Then there are dS minima for the range $104\lesssim \xi \lesssim 109$, 
with $ 60\lesssim   \sigma_{min} \lesssim 68$. If required, one can tune $\xi$ to obtain a Minkowski vacuum. One also finds that $\x <0.08 $ so one may consider the effect of $\alpha'$ corrections perturbatively.
To get a larger value of K\"ahler modulus, one can increase the number of D7-branes, as mentioned above. Choosing $b=0.02$,  corresponding to $n=314$ and with $A_{0}=1$ and $W_{0}=-8$ as above, 
the dS vacua exist for
$ 292\lesssim  \xi \lesssim 311$, corresponding to $  120 \lesssim \sigma_{min}  \lesssim145 $. 
Like before, $\x <0.08 $. A plot of the potential for $\xi=300$ is given in {\bf Fig \ref{VFxi}}.
We also note that the approximate relations given in Eq. (\ref{upapp}) are satisfied to a reasonable
accuracy in these two examples.


\section{The conifold}
\label{conifold}

We work in the background where the ten-dimensional metric  is in the form of warped geometry 
\be
    \exd s^2 = h(r)^{-1/2} \, \exd s^2_4 + h(r)^{1/2} \exd s^2_6 \,,
\ee
where $h(r)$ is the warp factor.

To incorporate the position of  D3-brane in the potential, we need to know $k(z,\ol{z})$, the K\"ahler potential  of  the CY space.
For applications to the throat we specialize to
the conifold defined by
\ba
\label{con}
\sum_{m=1}^{4} z_{m}^{2} =0 \, 
\ea
where $\{z_{m}, m=1,2,3,4\}$ are complex coordinates. 
The metric of the cone can also be written as
\ba 
 \exd s^2_6 =  \exd r^2 + r^2  ds_{T^{11}}^{2} 
 \ea
where $ds_{T^{11}}^{2}$ is the metric of the base, $T^{11}$, which is a five-dimensional Einstein manifold given by
\ba
ds_{T^{11}}^{2}= \left[ \frac19 \left( \exd \psi +
    \sum_{k=1,2} \cos\theta_k \, \exd \phi_k \right)^2 +
    \frac16 \sum_{k=1}^2 \Bigl( \exd \theta_k^2 + \sin^2\theta_k
    \exd \phi_k^2 \Bigr) \right]
\ea
where $\theta_{i} \in [0, \pi], \phi_{i} \in [0, 2 \pi]   $ and $\psi \in [0, 4 \pi]$ .
 
The metric of the cone is given by the K\"ahler potential 
\ba
\label{conk}
k= \left( \sum_{m=1}^{4} |z_{m} |^{2} \right)^{2/3} = r^{2} \, .
\ea

Eq. (\ref{con}) describes a singular cone. To remove the singularity at $r=0$, as in Klebanov-Strassler (KS)  \cite{Klebanov:2000hb}, one can work with the deformed conifold where the sum in Eq. (\ref{con}) 
is equal to $\epsilon^{2}$. Here $\epsilon^{2}$ is a parameter of dimension length to the power of 3 which
measures the deformation of the cone at the tip and its value is given by the value of warp factor at the tip. In this work, we assume that the D3-brane is located far away from the tip such that our analysis is
not sensitive to the details of the deformation. Furthermore, for simplicity we also assume that the infra-red geometry of the cone is cut-off at $r=r_{0}$. 

In the construction of KS, the warp factor is produced by the background $F_{3}$ and $H_{3}$ fluxes
turned on inside different three-cycles present inside the cone. To a good approximation, one can take 
the warp geometry in the form of 
\ba
h(r)= \frac{R^{4} }{r^{4} } \, ,
\ea 
where $R$ measures the curvature radius of the AdS geometry given by
\ba
R^{4} = \frac{27}{4}  \pi g_{s} N  \alpha'^{2}\, .
\ea
Here $N= MK$ is the background flux with $K$ and $M$ being the numbers of $H_{3}$ and $F_{3}$ fluxes respectively. Furthermore, the value of warp factor at the tip, $r=r_{0}$, denoted by $h_{0}$, is given by
\ba
h_{0} \equiv h^{-1/4}(r_{0})=  \frac{r_{0}}{R}=
e^{-2\pi K/3 g_{s}M} \, .
\ea

We are interested in the radial motion of the mobile D3-brane inside the throat. In the absence of
interaction between D3 and D7-branes, the D3-brane moves freely in the AdS-type background above.
However, once the interaction between D3 and D7-branes are taken into account, D3-brane will
not be free anymore. We would like to study the motion of the D3-brane 
once the perturbative K\"ahler corrections and nonperturbative correction to superpotential are included.
This problem was previously studied in \cite{BCDF, Baumann2, Krause:2007jk, Pajer:2008uy, Chen:2008ad} where the background is in the limit of KKLT where $|W_{0}| \ll 1$ by tuning and 
the perturbative corrections to K\"ahler potential is neglected, i.e. $\xi=0$. 

In  \cite{BCDF}
the motion of D3-brane in Ouyang embedding \cite{Ouyang:2003df} was studied. In \cite{Baumann2} this was generalized to other embeddings such as Kuperstein embedding \cite{Kuperstein:2004hy}.
In this work we formulate the potential for the D3-brane for generic embeddings and specialize to the above two embeddings.

From the K\"ahler  potential of the cone given as in Eq. (\ref{conk}), one obtains
\ba
k_{ i\bar j} = \frac{2}{ 3 r} \left[  \delta_{ i \bar j} + \frac{ z_{i} \bar z_{j} }{ |z_{4}|^{2}} 
-\frac{1}{  3 r^{3}} \left(  z_{i} \bar z_{j} + z_{j} \bar z_{i} - z_{i} z_{j} \frac{\bar z_{4}}{ z_{4}}
- \bar{ z_{i}}\bar{ z_{j} } \frac{ z_{4}}{\bar  z_{4}}
\right) \right] 
\ea
where $i=1,2,3$ and we eliminated $z_{4}$ in terms of $z_{1}, z_{2}$ and $z_{3}$ using the definition of cone Eq. (\ref{con}). For the inverse metric, denoted by  $k^{\bar i j}_{(z)}$, we obtain
\ba
\label{inkz}
k^{\bar i j}_{(z)} = \frac{3 r}{2} \left[  \delta^{\bar i j} + \frac{\bar z_{i} z_{j} }{ 2 r^{3} } 
- \frac{\bar z_{j} z_{i} }{  r^{3} } \right] \, .
\ea

Alternatively, one can work with the $w$-coordinate which is defined by the following linear relation
to the z-coordinate
\ba
\label{zw}
\left(
 \begin{array}{ccc}
   w_{1}  & w_{2} \\
   w_{3} &      w_{4}  \\    
    \end{array}%
   \right )  
   = 
 \frac{1}{\sqrt{2}}  \left(
 \begin{array}{ccc}
   z_{1}+ i z_{2}  & z_{1} -i z_{2} \\
   z_{3} + i z_{4} &     - z_{3} + i z_{4}  \\    
    \end{array}%
   \right )  \, .
\ea
For a review of the construction of the cone in $z$ and $w$ coordinates see
{\bf Appendix \ref{appB}}.

The K\"ahler metric in $w$-coordinate is now given by
\ba
k_{i \bar j} = \frac{2}{3 r} \left[  \delta_{i \bar j}  + \frac{c_{i}^{k} c_{\bar j}^{s} }{|w_{3}|^{2}} 
w_{k} \bar w_{s} - \frac{1}{3 r^{3}} (  \bar w_{i} w_{j} + \frac{\bar w_{4}  }{w_{3}} c_{i}^{k} w_{k} w_{j}
+ \frac{w_{4}}{\bar w_{3}} c_{j}^{s} \bar w_{s} \bar w_{i} + \frac{|w_{4}|^{2}}{|w_{3}|^{2}} c_{\bar j}^{s} 
c_{i}^{k} \bar w_{s}  w_{k}  ) \right] 
\ea
where the non-zero components of the matrix $c^{i}_{j}$ given as in \cite{Baumann2}
are $c^{1}_{2}= c^{2}_{1}=1$ and  $ c^{4}_{3}=-1$.  Calculating the inverse metric, denoted by $k^{\bar i j}_{(w)}$, one obtains
\ba
\label{inkw}
k^{\bar i j}_{(w)} = \frac{3r}{2} \left[   \delta^{\bar i j} + \frac{\bar w_{i} w_{j}}{  2 r^{3}}
- \frac{ c^{i'}_{i} c^{j'}_{j} w_{i'} \bar w_{j'}     }{ r^{3}}
\right]  .
\ea
One can also check that the following relations hold which would be helpful later on in calculating
the potential
\ba
\label{relations}
k^{i} \equiv k_{\bar j} k^{\bar j i}_{(z)}= \frac{3}{2} z^{i} \quad , \quad 
k^{i} k_{i} =k=r^{2} \, ,
\ea
with a similar  relations also for $w$-coordinate.


\section{D3-D7 Brane Inflation Dynamics}
\label{embeddings}
In this section we calculate the potential between D3 and D7 branes in the throat and see whether or not a period of slow-roll inflation can be achieved( for an update on D3-D7 inflation \cite{Dasgupta:2002ew}, see \cite{Haack}). The D3-D7 potential is a function of the radial as well as the azimuthal positions of the D3-brane. After finding the stable angular directions of the mobile brane, the potential is studied as a function of the radial position of the D3-brane which plays the role of the inflation field.

To calculate the F-term potential we need to specify the form of the holomorphic function $F(z)$, determining the embedding of the stack of D7-branes in this configuration. As explained before, we shall specialize to  Kuperstein and Ouyang embeddings which define two characteristic classes of embeddings.

\subsection{Ouyang Embedding}

The Ouyang embedding is defined in the $w$-coordinate via
\ba
\label{OuyF}
    F(w) &=& 1 - \frac{w_1}{\mu} \nonumber\\
    &=&1 - (\frac{r}{r_{\mu}})^{3/2}  \, e^{\frac{i}{2} (\psi - \phi_1 -
    \phi_2)} \sin \frac{\theta_1}{2} \sin \frac{\theta_2}{2} \,.
\ea
Here $\mu$ is a parameter of dimension length to the power of $3/2$. One can define the length parameter $r_{\mu}$ via $r_{\mu }= \mu^{2/3}$. Schematically, $r_{\mu}$ represent the distance of the
D7-brane from the throat. 
Using the embedding Eq. (\ref{OuyF}), the contribution of the D3-brane to the superpotential is 
\ba
A(w_{1})= A_{0} (1-   \frac{w_{1}}{\mu})^{1/n} \, .
\ea

Including the contribution of the D3-brane into superpotential, the F-term potential calculated from
Eq. (\ref{totalVF})  is now given by
\ba
\label{D3VFO}
V_{F}= \frac{\cF^{1/n}}{3 R \hR} \left[ a_{1} + a_{2} \left( 2- \frac{w_{1} + \bar w_{1}}{\mu}  \right) \cF^{-1}   
+ a_{3} \cF^{-1/2n} + a_{4} k^{\bar 1 1}_{w} \cF^{-1} \right]
+ \frac{a_{5}}{  3 R \hR   }
\ea
where
\ba
\label{cF}
\cF \equiv |F|^{2} = \left(  1-  \frac{w_{1} + \bar w_{1}}{\mu} +  \frac{|w_{1}|^{2}}{\mu^{2}} 
\right)
\ea
and $a_{i}$ are defined via
\ba
\label{ai}
a_{1}&=& |A_{0}|^{2} e^{-2 b \sigma} \left( (\bR +\gamma r^{2} ) b^{2} + 6 b \frac{\bR}{\hR} 
+ \frac{9}{\hR^{2}} ( \bR -\hR   )- {3b \over n}  \right)   \nonumber\\
a_{2}&=& { 3b \over 2n}  e^{-2 b \sigma}  |A_{0}|^{2} \nonumber\\
a_{3} &=&  6 W_{0} A_{0} e^{- b \sigma} \left(b {\bR \over \hR}  +  \frac{3}{\hR^{2}} ( \bR -\hR   ) \right)
\nonumber\\
a_{4}&=& \gamma^{-1} e^{-2 b \sigma} ({  A_{0}\over \mu \,  n  })^{2}  \nonumber\\
a_{5}&=&  \frac{9}{\hR^{2}} ( \bR -\hR   ) W_{0}^{2} \, .
\ea
Note that due to the particular form of embedding given by Eq. (\ref{kupF}), only the $({\bar 1 1})$ component of the inverse metric $k^{\bar i j}_{(w)}$ shows up in the F-term potential with
\ba
k^{\bar 1 1}_{(w)} &=& \frac{3 r}{2}   \left( 1+ \frac{|w_{1}|^{2}}{ 2 r^{3}  } - \frac{ |w_{2}|^{2}  }{ r^{3}  }   \right)
\nonumber\\
&=& \frac{3 r}{16}   \left(  7 - 3 \cos \theta_{1} - 3 \cos \theta_{2} - \cos \theta_{1} \cos \theta_{2} \right) \, .
\ea

The details of the stability analysis is relegated to {\bf Appendix
\ref{appC}}. Here we summarize the important
results. Possible stable solutions are given by
\ba
(a) : \theta_{1}= \theta_{2}=0 \quad  \quad  ~  \quad \quad  \quad  \quad   \quad  \quad   \quad \quad 
 (b) : \ps= 2 m \pi, \theta_{1}= \theta_{2}=\pi   
\ea
where $\ps \equiv \psi -\phi_{1} -\phi_{2}$ and $m$ is an integer  subject to $- 4 \pi \leq   \ps \leq 4\pi$.


\subsubsection{Stability of case (a)}

In case (a) we note that $\ps$ is a flat direction and the potential is independent of $\ps$.
Also $w_{1}=0, |w_{2}|=1$ which gives $F=\cF=1$.
In order for the solution to be stable, the following approximate inequality should be met
\ba
\label{stablea}
16 \pi^{2} \gamma^{2} \mu^{2} W_{0}^{2} e^{2 b \sigma} \,  r \cos^{2} \ps /2 <1 \, .
\ea

On the other hand, using the relations  (\ref{sigma}) and (\ref{gamma})  
and taking $r_{\mu} \lesssim R_{CY}$, one has
\ba
\label{gammamu}
\gamma  \, r_{\mu}^{2} \lesssim  (R_{CY} m_{s})^{4} \frac{m_{s}^{4}}{M_{P}^{2}} R_{CY}^{2} \sim 1
\ea
Using this and noting that $b \sigma \simeq 3$ as in Eq. (\ref{upapp}), then
 Eq. (\ref{stablea}) is satisfied if
\ba
\label{rrmu}
\frac{r}{r_{\mu}} \cos^{2} \ps  < 10^{-5} W_{0}^{-2} \, .
\ea
The consistency of our picture requires that $r > r_{0}$, where $r_{0}$ is the IR cut-off of the throat.
Assuming that $r_{\mu}$ is comparable to the AdS scale of the throat, $R$, then in order for Eq. (\ref{rrmu}) to be satisfied we need that, $h_{0}$, the warp  factor at the tip of the throat,
to be as small as $10^{-5} W_{0}^{-2}$. Otherwise, one has to tune  $\cos \ps/2$ sufficiently close to $0$. 

Having stabilized the angular directions, we can now look at the behavior of the potential
as a function of $r$.  Because $F=\cF=1$, as in \cite{BCDF},
we see that the contribution
of the mobile D3-brane to the superpotential vanishes and $A=A_{0}$. This is subsequently
labeled as ``delta-flat'' solution in \cite{Baumann2}.  Here ``delta'' represents the change into $V_{F}$ due to contribution from D3-brane into superpotential.

One may ask whether or not a period of slow-roll inflation is supported while the brane is moving
towards the tip. With $\cF=1$, we obtain
\ba
\label{VFOuyang}
V_{F} = \frac{1}{3 R \hR} (a_{1}  + 2 a_{2}+  a_{3} + a_{5})  
\simeq V_{0} \left[1+ \frac{\gamma}{ \sigma } r^{2} \left( 1+ \frac{5 a_{5} }{4(a_{1}+a_{3}+ a_{5} )} \right) \right]
\ea
where $V_{0}$ represents the value of potential in the absence of D3-brane, corresponding to setting
$r=0$. To calculate the slow-roll parameter $\eta$, first we find the normalized field, $\phi$,
given by
\ba
\phi = \sqrt{ \frac{3 \gamma}{\sigma} }\, r
\ea
where the brane kinetic energy has the standard form $-\partial_{\mu} \phi   \partial^{\mu} \phi$/2. 
Calculating $\eta$ one obtains that 
\ba
\eta =  \frac{\sigma}{3 \gamma}  \left| \frac{V,rr}{V}  \right|  
  \simeq \frac{2}{3}  + \frac{5 a_{5} }{6 (a_{1} +a_{3}+ a_{5} )}  \, .
\ea
Interestingly enough, the first term containing the factor of $2/3$ 
is similar to the result of \cite{KKLMMT}. The second term containing $a_{5}$ is due to
$\xi$ corrections to the K\"ahler potential where the no-scale property of the four-dimensional susy
is broken. On the other hand,  $a_{5}>0$ and $a_{1}+ a_{3}+ a_{5} \sim - a_{2}$ with $a_{2}\ll a_{5}$ in our background. This indicates that $\eta >1$ and
it is not possible to obtain slow-roll inflation in this embedding without fine-tuning.

\begin{figure}[t] 
\vspace{-0.8cm}
   \centering
   \includegraphics[width=3in]{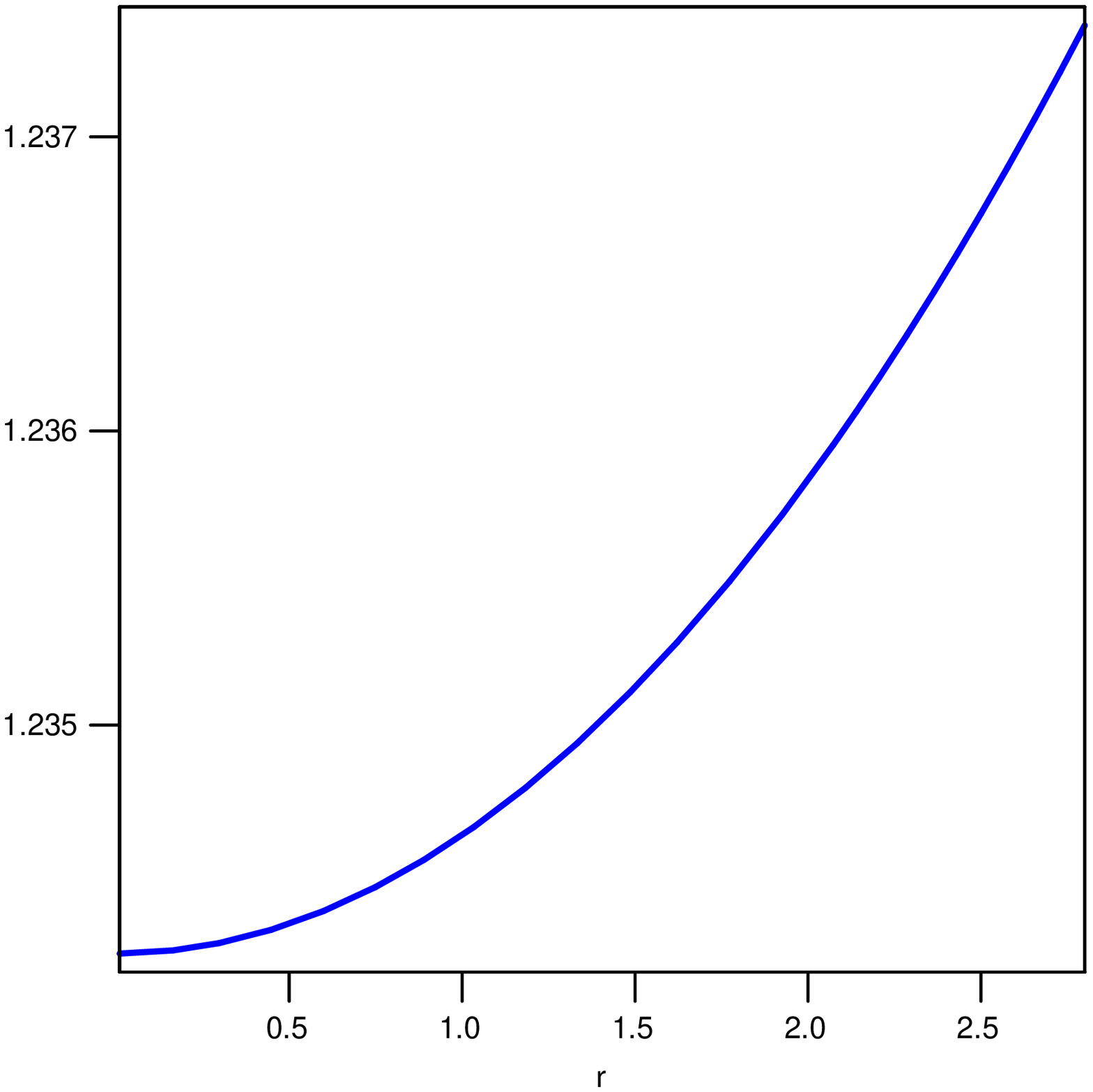}  
\includegraphics[width=3in]{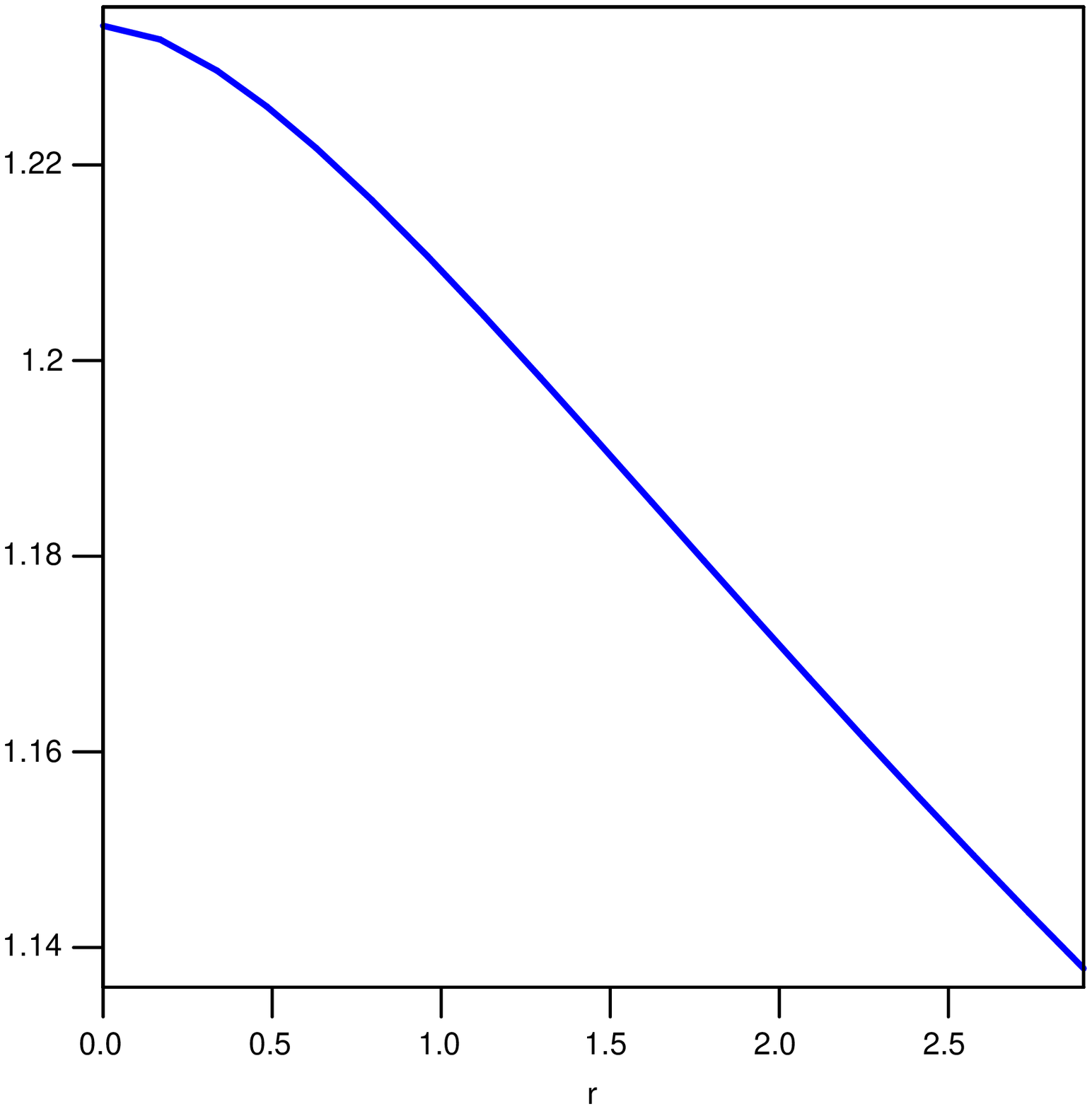} 
   \caption{In this figure, $V_{F}$(in units of $10^{-8}$) is plotted as a function of $r$ 
   for $A_{0}=1, W_{0}=-8, n=314$ with $\xi=300$, $\gamma=10^{-3}$ and $r_{\mu}=2.9 $ in Ouyang embedding. The left figure is for case (a) where the D3-brane is attracted towards the tip while the right figure is for case (b) where the D3-brane is attracted towards the stack of D7-branes. }
\label{Vrab}
\end{figure}

The figure on the left hand side of {\bf Fig. (\ref{Vrab})} represents 
$V_{F}$ for case (a). We see that as in \cite{BCDF}, the D3-brane is attracted towards
the tip of the throat. This is also easily seen from Eq. (\ref{VFOuyang}).


\subsubsection{Stability of case (b)}

In order for case (b) to be a stable solution, we require that  $\cos \ps /2 =-1$ and $\ps=2 \pi$. This corresponds to $w_{1} = -r^{3/2}$. Furthermore, the position of the D3-brane should satisfy
the condition $r_{c}< r<r_{\mu}$ where
$r_{c}$ is approximately given by
\ba
\label{rc}
r_{c} \simeq ( 8 \pi \gamma \, \mu \, W_{0} e^{b \sigma} )^{-2} \, .
\ea
For, $r_{0}\leq r \leq r_{c}$ the solution is unstable.

In this example we see that the contribution of D3-brane to the superpotential does not vanish, corresponding to 
$F  =1 +( r/r_{\mu})^{3/2}$, so we find a specific ``non-delta flat`` solution in Ouyang embedding. In the work of
\cite{BCDF} this solution was not a stable solution. However, in our case here, the stability
of the solution is due to the additional term containing $a_{3}$. With
$|W_{0}| \sim 1$, the term containing $a_{3}$ dominates over other terms and plays the key role
in making case (b) to be a stable solution. 

To calculate $\eta$, we note that
\ba
\label{VFKup}
V_{F} \simeq   \frac{1}{3 R \hR} \left(  a_{1}+ a_{3} \cF^{1/2n} + a_{5} \right) \, .
\ea
Calculating $\eta$ one obtains
\ba
\label{etbO}
\eta & \simeq& \frac{2}{3}  + \frac{5 a_{5} }{6 (a_{1} +a_{3}+ a_{5} )} +
 \frac{\sigma \,  a_{3} }{ 4n \gamma \mu r^{1/2} F^{2} \, (a_{1}+ a_{3}+ a_{5})   }  
\left( 1- \frac{2 r^{3/2}}{\mu}   \right )  \nonumber\\
&\simeq& \frac{\sigma \,  a_{3} }{ 4n \gamma \mu r^{1/2}  F^{2}\, (a_{1}+ a_{3}+ a_{5})   }  
\left( 1- \frac{2 r^{3/2}}{\mu}   \right ) \, .
\ea
Looking at the behavior of $\eta$, one observes that the potential has an inflection point
at $r_{in}$ given approximately  by
\ba
r_{in} \simeq  2^{-2/3} r_{\mu}\, 
\ea
where $\eta=0$. However, the potential does not stay nearly flat near the inflection point.
As in case (a), the potential is too steep to allow inflation. 

One also is interested in the shape of potential and whether or not the D3-brane is moving towards the tip or not.  In {\bf Fig(\ref{Vrab})} we have plotted the potential as a function of $r$ 
where $\sigma$
is stabilized to to its background value. We see that the D3-brane is attracted towards the D7-brane.
This can also be understood analytically. The minimum, $r_{min}$, of the potential (\ref{VFKup}) 
is determined by
\ba
(\frac{r_{min}}{r_{\mu}} )^{1/2} \left(   1+ (\frac{r_{min}}{r_{\mu}} )^{3/2} \right) 
&\simeq &
\frac{  3\,  \sigma   }{   2 n\,  \gamma r_{\mu}^{2}  } \frac{ - a_{3}  }{  a_{3} + \frac{9}{4} a_{5}   } \nonumber\\
&\simeq&  \frac{   1  }{   \gamma r_{\mu}^{2}  }  \gtrsim 1 \, .
\ea
To go to the second line,  we used the approximate relation $b \sigma \sim 3$, equivalent to
$\sigma \sim 3n/2\pi$, which holds in our background. The final expression is also a consequence
of Eq. (\ref{gammamu}).  In conclusion
\ba
\label{rminb}
\frac{r_{min}}{r_{\mu}} \gtrsim 1\, .
\ea
This indicates that  the minimum of the D3-position is typically outside of the throat. The D3-brane is moving from the IR region into the UV region and get dissolved into D7-branes.
This may provide an explicit construction of models of IR brane inflation\cite{Chen:2004gc}
where the D3-brane leaves the IR region towards the UV region of the throat.  
However, one may tune the background parameters such as $n$ and $W_{0}$
such that $r_{min}$ is close to the the position of
the D7-brane but inside the throat. In this case the D3-brane is
confined between the tip of the throat and the D7-branes.


\subsection{Kuperstein Embedding}

The Kuperstein embedding is define by 
\ba
\label{kupF}
F= F(z_{1})= 1- \frac{z_{1}}{\mu} \, .
\ea

The F-term potential is formally the same as in the Ouyang embedding, with the replacement of
$w_{1} \rightarrow z_{1}$ and $k^{\bar 1 1}_{(w)} \rightarrow k^{\bar 1 1}_{(z)}$ where
\ba
k^{\bar 1 1}_{(z)} = \frac{3r}{2} \left( 1- \frac{|z_{1}|^{2} }{ 2 r^{3}}    \right) \, .
\ea

The stable solutions are 
\ba
(a) : && \theta_{1}= \theta_{2}=0, \, \varphi+\tilde{\phi}=2n\pi \nonumber\\ 
(b) :&& \theta_{1}= \theta_{2}=\pi, \, \varphi-\tilde{\phi}=2n\pi\nonumber\\
(c) :  && \theta_1=\theta_2, \, \varphi=(2n+1)\pi, \, \tilde{\phi}=(2m+1)\pi
\ea
where $m$  and $n$ are integers and $\tilde \phi \equiv \phi_{1}+ \phi_{2}$.

In {\bf Apendix C} we show that only solutions that lead to $z_1=-\frac{r^{3/2}}{\sqrt{2}}$ are stable 
\cite{Baumann2}.  This requires that $n$ to be odd(even) in case (a)$\left( (b) \right)$, while
$m+n$ to be even in case (c).

For all three cases, we note that $F= 1+ r^{3/2}/\sqrt 2 r_{\mu}^{3/2}$ and $V_{F}$ is formally given as in Eq. (\ref{VFKup}). To calculate $\eta$, we borrow Eq. (\ref{etbO}) with the replacement $\mu \rightarrow \sqrt 2 \mu$  which 
gives
\ba
\eta \simeq  \frac{ \sigma \,  a_{3} }{ 4n \sqrt 2 \gamma \mu r^{1/2}  F^{2}\, (a_{1}+ a_{3}+ a_{5})   }  
\left( 1- \frac{\sqrt 2 r^{3/2}}{\mu}   \right ) \, .
\ea
As  before, for parameters of physical interest, this gives too big a value of $\eta$ to allow for slow-roll
inflation. Furthermore, there is inflection point at
$$r_{in } \simeq  2^{-1/3} r_{\mu}, $$
but the potential does not stay flat enough around the inflection point to allow slow-roll inflation as 
is clear from the large value of $\eta$ given above.

The shape of the potential is similar to case (b) of the Ouyang embedding; the D3-brane is attracted towards the D7-brane if $r_{min}> r_{\mu}$. However, as before, one can also arrange such that
$r_{min} \lesssim r_{\mu}$ and the D3-brane is confined between the tip of the throat and the D7-branes.

\section{Conclusion and Discussions}
\label{conclusions}

In this work the dynamics of D3-D7 branes in an inflationary throat was studied.
In the limit where $W_{0}$ is large and $\alpha'$ corrections to K\"ahler potential are included,
it is shown that the dS uplifting and the moduli stabilization can be achieved without
the addition of anti-D3 branes. This has the advantage that the SUSY is not broken explicitly.

We have treated $ \xi $, the parameter corresponding to $\alpha'$ correction, as a
constant. In a more realistic treatment, the $\alpha'$ correction is a function of the dilaton field. We work in a background where the dilaton is fixed. However, in calculating the covariant derivatives
of $W$ and $K$, one should consider dilaton as an independent field and only at the end impose the condition
that dilaton is constant. This brings corrections of order $\xi^{2}$ into our background potential
Eq. (\ref{VF1}). Consequently, this will change the numerics of the uplifting slightly
without changing the whole picture. Furthermore, 
this also does not affect our conclusion in section 4 about the stability
analysis of different embeddings studied. The reason is that the terms containing
$\xi$ are only important in $a_{5}$ and our discussions rely on the leading term 
in $a_{5}$ which is linear in $\xi$.
Therefore, in order to keep the analysis transparent and tractable, we have neglected the effects of dilaton in $\xi$. 

In this work, we have considered the leading $\alpha'$ correction to  K\"ahler potential. 
We work in the large volume limit where the higher order $\alpha'$ corrections, which are unknown, may be neglected. It would be interesting to see how higher order corrections to K\"ahler potential may affect the uplifting mechanism. However, it is expected that our stability analysis remain unchanged, since our analysis there rely on the linear term in $\xi$ as explained above.

We have looked into the shape of the inflationary potential for the radial position of the D3-brane once all angular directions are integrated out. The analysis were specifically performed for Ouyang and Kuperstein embeddings which define two characteristic  classes of embeddings for the wrapped D7-branes. In  Kuperstein embedding the D3-brane is attracted
towards the stack of D7-branes while for Ouyang embedding it can move towards the tip
of the throat or towards the D7-branes, depending on the stabilized angular directions.
We have seen that the potential in both cases is too steep to allow
a period of slow-roll inflation. This has origin in largeness of $|W_{0}|$ which is required in our
uplifting mechanism. Unlike \cite{Baumann2}, the contributions to the potential from 
nonperturbative corrections to the superpotential  can not be made large enough to cancel out the contributions from $W_{0}$.
It is expected that this conclusion to be generic, holding for other embeddings too. It would be interesting to perform a systematic investigation for other embeddings to see whether or not this expectation still holds.

\vspace{1cm}
\section*{Acknowledgments}

We thank Cliff Burgess, Jim Cline and Keshav Dasgupta for their initial contributions and for many discussions and comments. We Also thank X. Chen for discussions. H.F. thanks Banff International Research Station (BIRS) for hospitality when this work was initiated.
\appendix

\section{  Uplifting by $\alpha'$ correction  }
\label{appA}
Here we study uplifting and K\"ahler modulus stabilization is more details.
The potential, up to second order in $\xi$, is given by Eq. (\ref{upapp})
$$
\label{VFxi2}
V_{F} \simeq \frac{b W_{0}}{ 2 \sigma^{2}}   e^{- b \sigma} 
+ \frac{3 W_{0}^{2} \xi }{  2^{11/2}  \,  \sigma^{9/2}  } 
- \frac{9 W_{0}^{2} \xi^{2} }{  256\,  \sigma^{6} }  \, .
$$
As mentioned, we have neglected terms containing $e^{-2 b \sigma}$, which is a good approximation
for large enough $W_{0}$. The minimum of  Eq. (\ref{upapp}) is given by
\ba
\label{f}
|W_{0} | \xi b^{3/2} =  \frac{ 2^{11/2} x^{5/2} (x+2)  e^{-x} }{ 27\left(1 -  2^{-1/2}\,  \xi b^{3/2} x^{-3/2} \right) }  
\equiv f(x)
\ea
where we have defined $b \sigma = x$.
Furthermore, requiring that the potential is positive at the minimum, implies that
\ba
\label{g}
|W_{0} | \xi b^{3/2} >    \frac{  2^{9/2}\,  e^{-x} x^{5/2}   }{3 \left( 1 - \frac{3}{2^{5/2}} \xi b^{3/2} x^{-3/2} \right) } \equiv g(x)
\ea

With some efforts, one can check that the functions $f$ and $g$ have maximum at $x_{f}$ and $x_{g}$,
respectively, given by
\ba
\label{xfg} x_{f} \simeq 3.1 - 3.2\,  \x      \quad , \quad 
x_{g} \simeq 2.5 - 2.3 \,  \x
\ea

Looking at the plots of $f$ and $g$, one concludes that in order to satisfy conditions 
(\ref{f}) and (\ref{g}), one requires that    $x_{g} \leq x \leq x_{f}$ and 
$f(x_{g}) \leq   |W_{0}|^{2} \xi b^{3/2}     \leq f(x_{f})$, or
\ba
\label{xbound}
2.5 -2.3 \,   \x \leq b \sigma \leq 3.1 - 3.2 \, \x \quad , \quad
6.1 + 9.2 \,  \x \leq   |W_{0}|^{2} \xi b^{3/2}   \leq 6.5+  13\,   \x \, .
\ea
This leads to the following useful relations
\ba
\label{upapp2}
b \sigma \simeq 3  \quad , \quad | W_{0}| \xi b^{3/2} \simeq 6.5 \, .
\ea
We have checked many numerical examples where the relations in Eq. (\ref{upapp2})
are hold with reasonable accuracies.

\section{The Conifold}
\label{appB}

Here we briefly summarized the representation of conifold in $w$ and $z$ coordinates.
The conifold is defined by 
\ba
\sum_{m=1}^{4} z_{m}^{2} =0 \, 
\ea
with the K\"ahler potential 
\ba
k= \left( \sum_{m=1}^{4} |z_{m} |^{2} \right)^{2/3} = r^{2} \, .
\ea

The conifold in $z$-coordinate is given by
\ba
z_{1}&=& \frac{r^{3/2}}{\sqrt 2}  e^{i \psi/2} \left[ \cos \left(  \frac{\theta_{1}+ \theta_{2}}{2}  \right)
\cos \left(  \frac{\phi_{1}+ \phi_{2}}{2}  \right) + i \cos \left(  \frac{\theta_{1}- \theta_{2}}{2}  \right)
\sin \left(  \frac{\phi_{1}+ \phi_{2}}{2}  \right) \right]
\nonumber\\
z_{2}&=& \frac{r^{3/2}}{\sqrt 2}  e^{i \psi/2} \left[ -\cos \left(  \frac{\theta_{1}+ \theta_{2}}{2}  \right)
\sin \left(  \frac{\phi_{1}+ \phi_{2}}{2}  \right) + i \cos \left(  \frac{\theta_{1}- \theta_{2}}{2}  \right)
\cos \left(  \frac{\phi_{1}+ \phi_{2}}{2}  \right) \right] \nonumber\\
z_{3}&=& \frac{r^{3/2}}{\sqrt 2}  e^{i \psi/2} \left[ -\sin \left(  \frac{\theta_{1}+ \theta_{2}}{2}  \right)
\cos \left(  \frac{\phi_{1}- \phi_{2}}{2}  \right) + i \sin \left(  \frac{\theta_{1}- \theta_{2}}{2}  \right)
\sin \left(  \frac{\phi_{1}- \phi_{2}}{2}  \right) \right] \nonumber\\
z_{4}&=& \frac{r^{3/2}}{\sqrt 2}  e^{i \psi/2} \left[ -\sin \left(  \frac{\theta_{1}+ \theta_{2}}{2}  \right)
\sin \left(  \frac{\phi_{1}- \phi_{2}}{2}  \right) - i \sin \left(  \frac{\theta_{1}- \theta_{2}}{2}  \right)
\cos \left(  \frac{\phi_{1}- \phi_{2}}{2}  \right) \right]  .
\ea
Notice that $r$ denotes proper distance up the throat, and $r = 0$
corresponds to the throat's tip, which is singular if the conifold is not deformed.

Upon linear transformations (\ref{zw}) one goes to the $w$-coordinate where now the conifold is
given by
\bea
    w_1 &=& r^{3/2} e^{\frac{i}{2} (\psi - \phi_1 - \phi_2)} \sin
    \frac{\theta_1}{2} \sin \frac{\theta_2}{2} \nn\\
    w_2 &=& r^{3/2} e^{\frac{i}{2} (\psi + \phi_1 + \phi_2)} \cos
    \frac{\theta_1}{2} \cos \frac{\theta_2}{2} \\
    w_3 &=& r^{3/2} e^{\frac{i}{2} (\psi + \phi_1 - \phi_2)} \cos
    \frac{\theta_1}{2} \sin \frac{\theta_2}{2} \nn\\
    w_4 &=& r^{3/2} e^{\frac{i}{2} (\psi - \phi_1 + \phi_2)} \sin
    \frac{\theta_1}{2} \cos \frac{\theta_2}{2} \,. \nn
\eea
%


\section{The stability analysis}
\label{appC}

Here we study in more details the angular stability of different solutions for both Ouyang and Kuperstein embeddings

\subsection{Ouyang embedding}

The independent angular variables  are $\{ \Psi _{i} \} = \{ \theta_{1}, \theta_{2} , \ps \}$, 
where $\ps \equiv \psi- \phi_{1} -\phi_{2}$.
Furthermore,  we note that $V_{F}$ as given in Eq. (\ref{D3VFO}) is a function of $|w_{1}|^{2}, |w_{2}|^{2}$ and $w_{1} + \bar w_{1}$. Define $x_{1}= |w_{1}|^{2},  x_{2}= |w_{2}|^{2}$ 
and $x_{3}= w_{1} + \bar w_{1} $.
As argued in \cite{Baumann2},  finding the stable values of $\Psi_{i}$ independent
of $r$ requires that  
\ba
\label{exO}
\frac{\partial x_{i}}{ \partial \Psi_{j}  } =0
\ea
The sets of solutions to above equations are 
\ba
(a) : && \theta_{1}= \theta_{2}=0 \quad  \quad  ~  \quad \quad  \quad  \quad   \quad  \quad   \quad \quad  (b) : \ps= 2 m \pi, \theta_{1}= \theta_{2}=\pi   
\nonumber\\ 
(c) :  &&  \ps= (2 m+1) \pi, \theta_{1}= 0,\theta_{2}=\pi    \quad  \quad 
(d) :  \ps= (2 m+1) \pi, \theta_{1}= \pi,\theta_{2}=0  \, ,
\ea
where $m$ is an integer  subject to $- 4 \pi \leq   \ps \leq 4\pi$.

Below we demonstrate that only cases (a) and  (b) could be stable.
Define $X_{i} = \partial V /\partial x_{i}  $. Then at the extrema given by Eq. (\ref{exO}) the 
mass matrix, $V_{ij}$, is given by
\ba
\label{Vij}
V_{ij} = \frac{  \partial^{2} V   }{ \partial \Psi_{i} \partial \Psi_{j} }
= X_{k} \frac{ \partial^{2} x_{k}}{ \partial \Psi_{i} \partial \Psi_{j}   } \, .
\ea 
The nonzero components of the mass matrix  are:
\ba
\label{Xall}
(a)&&  \quad \quad   V_{\theta_{1} \theta_{1}} = V_{\theta_{2} \theta_{2}} = -  \frac{r^{3}}{2} X_{2}
\quad , \quad V_{\theta_{1} \theta_{2}} =   \frac{r^{3/2}}{2}  X_{3}  \cos \frac{\ps}{2}  \nonumber\\
(b)&& \quad \quad   V_{\theta_{1} \theta_{1}} = V_{\theta_{2} \theta_{2}} = -  \frac{r^{3}}{2} X_{1}
- \frac{r^{3/2}}{2} \cos \frac{\ps}{2}     X_{3}
\quad , \quad V_{\ps \ps} =  - \frac{r^{3/2}}{2} \cos \frac{\ps}{2}     X_{3}
 \nonumber\\
 (c)&&  \quad \quad   V_{\theta_{1} \theta_{1}} = \frac{r^{3}}{2} X_{1} \quad , \quad
 V_{\theta_{2} \theta_{2}} =   \frac{r^{3}}{2} X_{2}
 \quad , \quad V_{\theta_{1} \ps} =  - \frac{r^{3/2}}{2} \sin \frac{\ps}{2}    X_{3} \nonumber\\
 (d)&&  \quad \quad   V_{\theta_{1} \theta_{1}} = \frac{r^{3}}{2} X_{2} \quad , \quad
 V_{\theta_{2} \theta_{2}} =  \frac{r^{3}}{2} X_{1}
 \quad , \quad V_{\theta_{2} \ps} =  - \frac{r^{3/2}}{2} \sin \frac{\ps}{2}    X_{3}
\ea 

In the cases of (c) and (d) one can check that there is a negative eigenvalue, so these cases are
unstable, corresponding to $w_{1}= w_{2} =0$. 
The stability of the other two solutions depends on the sign of $X_{i}$ given by
\ba
\label{X13}
X_{2} &=& - \frac{ a_{4} \,  \cF^{\frac{1}{n} -1} }{ 2 r^{2}\,  R \hR   }  \nonumber\\
X_{1}&=& \frac{  \cF^{ \frac{1}{n} -1  }  }{3 \mu^{2} R \hat R} \left[  \frac{a_{1} }{n} 
-   a_{2} ( 1- \frac{1}{n}   )  (   2- \frac{w_{1} + \bar w_{1}}{\mu}  ) \cF^{-1}
+ \frac{a_{3}}{2 n } \cF^{  \frac{-1}{2 n}  }  + \frac{ 3 \mu^{2}   }{ 4 r^{2}  }  a_{4} 
- a_{4} (  1- \frac{1}{n} ) k^{\bar 1 1 }_{w} \cF^{-1}
\right]  \nonumber\\
X_{3}&=& \frac{  - \cF^{ \frac{1}{n} -1  }  }{3 \mu R \hat R} \left[  \frac{a_{1} }{n} + a_{2}
- a_{2} ( 1- \frac{1}{n}   )  (   2- \frac{w_{1} + \bar w_{1}}{\mu}  ) \cF^{-1} + 
 \frac{a_{3}}{2 n } \cF^{  \frac{-1}{2 n}  } 
- a_{4} (  1- \frac{1}{n} ) k^{\bar 1 1 }_{w} \cF^{-1} 
\right]  \nonumber\\
\ea

From the form of $a_{i}$ given in Eq. (\ref{ai}) we note that all $a_{i}$ except $a_{3}$ are positive.
This implies that $X_{2}<0$. 
Furthermore, $|a_{3}|/n > a_{1}/n > a_{2}$ which will be useful later.


\subsubsection{Stability of case (a)}

For case (a) we note that $\ps$ is a flat direction and the potential is independent of $\ps$.
In order for solution to be stable we require that $X_{2}<0$ and 
$r^{3} X_{2}^{2} - X_{3}^{2}    \cos^{2} \ps /2 >0$. As mentioned above, the first condition is always  satisfied. On the other hand, we have $w_{1}=0, |w_{2}|=1$ which gives $k^{\bar 1 1}_{w}=0$ and
$\cF=1$. Calculating $X_{3}$, we find 
\ba
\label{X3ap}
X_{3}= \frac{  - 1  }{3 \mu R \hat R} \left(  \frac{a_{1}}{n} + ( \frac{2}{n} -1) a_{2} + \frac{a_{3}}{2 n} 
\right) \simeq 
 \frac{  - a_{3}  }{6 n\,  \mu R \hat R} \, .
\ea
The stability of the solution is therefore approximately translated into
\ba
16 \pi^{2} \gamma^{2} \mu^{2} W_{0}^{2} e^{2 b \sigma} \,  r \cos^{2} \ps /2 <1 \, .
\ea


\subsubsection{Stability of case (b)}

For the stability of case (b) we require that $  X_{3} \cos \ps /2 <0$ and 
$X_{3} \cos \ps /2 + r^{3/2} X_{1} <0$. 
Furthermore
\ba
\label{X3b}
X_{3} \sim - \frac{\cF^{-1+ \frac{1}{n}}}{ 3 \mu R \hR} \frac{a_{3}}{2n} >0 \, .
\ea
Demanding that $  X_{3} \cos \ps /2 <0$ requires that $\cos \ps /2 =-1$ and $\ps=2 \pi$. This corresponds to $w_{1} = -r^{3/2}$.

On the other hand, one can also check that
\ba
\label{X13b}
X_{3} \cos \ps /2 + r^{3/2} X_{1}  \sim  \frac{\cF^{-1+ \frac{1}{n}}}{ 3 \mu^{2} R \hR}
\left[  \frac{a_{3}}{2n}  \mu \cF^{-1/2n} ( 1+ \frac{r^{3/2}}{r_{\mu}^{3/2}} ) + \frac{3 \mu^{2}}{ 4 r^{1/2}} a_{4}
\right] \, .
\ea
Since $a_{3}<0$, the above expression changes sign for sufficiently small $r$.  In conclusion, we see that case (b) with $\ps = 2\pi$
is stable for the range $r_{c}< r<r_{\mu}$, while unstable for   $r_{0} <  r<r_{c}$, where
$r_{c}$ is approximately given by the root of the term in the bracket (\ref{X13b}) as
\ba
r_{c} \simeq ( 8 \pi \gamma \, \mu \, W_{0} e^{b \sigma} )^{-2} \, .
\ea


\subsection{ Kuperstein Embedding  }

Here we find the stable position of the angular variables $\{\Psi_i\}=\{\theta_1,\theta_2, \psi, \tilde{\phi}\}$,
where $\tilde{\phi}\equiv \phi_1+\phi_2$. The potential 
$V_F$ is is now a function of $z_1+\bar{z_1}$ and $|z_1|^2$. As in Ouyang embedding, 
the stable values of $\Psi_i$ satisfy:
\bea
\frac{\partial (z_1+\bar{z_1})}{\partial \Psi_j}=0,\:\:\:\frac{\partial |z_1|^2}{\partial \Psi_j}=0
\eea
The solutions are
\ba
(a) : && \theta_{1}= \theta_{2}=0, \, \varphi+\tilde{\phi}=2n\pi \nonumber\\ 
(b) :&& \theta_{1}= \theta_{2}=\pi, \, \varphi-\tilde{\phi}=2n\pi\nonumber\\
(c) :  && \theta_1=\theta_2, \, \varphi=(2n+1)\pi, \, \tilde{\phi}=(2m+1)\pi \nonumber\\
(d) :  &&  \theta_{1(2)}= 0,\, \theta_{2(1)}=\pi, \, \varphi=(2n+1)\pi, \, \tilde{\phi}=2m\pi    \nonumber\\
(e) : &&  \theta_{1(2)}= 0,\, \theta_{2(1)}=\pi,  \, \varphi=2n\pi, \, \tilde{\phi}=(2m+1)\pi
\ea
where $m$  and $n$ are integers.

Below we show that only solutions that lead to $z_1=-\frac{r^{3/2}}{\sqrt{2}}$ are stable 
\cite{Baumann2}. The mass matrix is given by 
\bea
V_{ij}=\frac{\partial V}{\partial\Psi_i\partial\Psi_j}=
Y_{1} \,  \frac{\partial^{2} (z_1+\bar{z}_1)}{\partial \Psi_i\Psi_j}
+Y_{2}\, \frac{\partial^{2} |z_1|^2}{\partial \Psi_i\Psi_j}
\eea
where 
\ba
Y_1\equiv \frac{\partial V_F}{\partial (z_1+\bar{z}_1)}
\quad , \quad 
Y2\equiv \frac{\partial V_F}{\partial |z_1|^2}
\ea
given by
\ba
Y_1&=&-\frac{\mathcal{F}^{\frac{1}{n}-1}}{3\mu R\hat{R}}
\left[\frac{a_1}{n}+ a_{2}-a_2(1-\frac{1}{n})(2-\frac{z_1+\bar{z_1}}{\mu})\mathcal{F}^{-1}
+\frac{a_3}{2n}\mathcal{F}^{\frac{-1}{2n}}-a_4(1-\frac{1}{n})k^{\bar{1}1}_z\mathcal{F}^{-1}\right]\nonumber\\
Y2&=&\frac{\mathcal{F}^{\frac{1}{n}-1}}{3\mu^2 R\hat{R}}
\left[\frac{a_1}{n}-a_2(1-\frac{1}{n})(2-\frac{z_1+\bar{z_1}}{\mu})\mathcal{F}^{-1}
+\frac{a_3}{2n}\mathcal{F}^{\frac{-1}{2n}}-a_4(1-\frac{1}{n})k^{\bar{1}1}_z\mathcal{F}^{-1}
-a_4\frac{3\mu^2}{4r^2}\right] \, . \nonumber\\
\ea
We see that $Y_{1}$ is formally the same as $X_{1}$ with the replacement of $w \rightarrow z$, while
$Y_{2}$ is similar to $X_{1}$ except with a change of sign in the term containing $\mu^{2} a_{4}$.
Since $|a_{3}|/n>a_{1}/n>a_{2}$ and $a_{3}<0$, $Y_1$ 
is always positive while $Y_2$ is always negative.

The nonzero components of the mass matrix are given by
\ba
(a)  &&V_{\theta_1\theta_1}=V_{\theta_2\theta_2}=-(-1)^{n}\frac{\sqrt{2}r^{3/2}}{4}Y_1-\frac{r^3}{4}Y_2 \, ,
\quad V_{\theta_{1} \theta_{2}} =  \cos \psi V_{\theta_1\theta_1} \nonumber\\
&&V_{\psi\psi}=V_{\tilde{\phi}\tilde{\phi}}=V_{\psi\tilde{\phi}}=-(-1)^{n}\frac{\sqrt{2}r^{3/2}}{4}Y_1 \, . \nonumber\\
(b)&& V_{\theta_1\theta_1}=V_{\theta_2\theta_2}=(-1)^{n}\frac{\sqrt{2}r^{3/2}}{4}Y_1-\frac{r^3}{4}Y_2 \, ,
\quad V_{\theta_{1} \theta_{2}} =  \cos \psi V_{\theta_1\theta_1} \nonumber\\
&&V_{\psi\psi}=V_{\tilde{\phi}\tilde{\phi}}=V_{\psi\tilde{\phi}}=(-1)^{n}\frac{\sqrt{2}r^{3/2}}{4}Y_1 \, . \nonumber\\
(c) && V_{\theta_1\theta_1}=V_{\theta_2\theta_2}=- V_{\theta_1\theta_2}=
(-1)^{m+n}\frac{\sqrt{2}r^{3/2}}{4}Y_1-\frac{r^3}{4}Y_2 \, ,
 \nonumber\\
&&V_{\psi\psi}= (-1)^{m+n}\frac{\sqrt{2}r^{3/2}}{4}Y_1 \, , \quad V_{\psi\tilde{\phi}} = \cos \theta_{1} \, V_{\psi\psi} \, \nonumber\\
&&V_{\tilde{\phi}\tilde{\phi}}= (-1)^{m+n}\frac{\sqrt{2}r^{3/2}}{4}Y_1 - \frac{r^{3}}{4} Y_{2} \sin^{2} \theta_{1} \, \, \nonumber\\
(d) && V_{\theta_1\theta_1}=V_{\theta_2\theta_2}= V_{\theta_{1} \theta_{2}} =\frac{r^3}{4}Y_2, \,  \quad
 \nonumber\\
&& V_{\theta_{1} \ps}= V_{\theta_{1} \tilde \phi}= V_{\theta_{2} \ps}= V_{\theta_{2} \tilde \phi}=
\frac{(-1)^{n+m} \sqrt 2}{4} r^{3/2}Y_1  \, . \nonumber\\
(e) && V_{\theta_1\theta_1}=V_{\theta_2\theta_2}= -  V_{\theta_1\theta_2}=
\frac{r^3}{4}Y_2, \, , \nonumber\\
&& V_{\theta_{1} \ps}= V_{\theta_{1} \tilde \phi}= -V_{\theta_{2} \ps}= V_{\theta_{2} \tilde \phi}=
\frac{(-1)^{n+m} \sqrt 2}{4} r^{3/2}Y_1  \, . 
\ea
  
Looking at the eigenvalues of the mass matrix, one easily finds that case (d) and (e) are unstable. 
More explicitly, one encounters the eigenvalues 
$$
\frac{r}{4} \left( r^{2} Y_{2} \pm (-1)^{m} \sqrt{  r^{4} Y_{2}^{2}  + 4 r Y_{1}^{2}       }
\right)
$$
which indicates the existence of instability.

The nonzero eigenvalues of case (a) are
\ba
\lambda^{a}_{1}= - \frac{(-1)^{n} \sqrt 2}{2} r^{3/2} Y_{1} \quad , \quad
\lambda^{a}_{2,3} = -\frac{1}{4} \left(  (-1)^{n} \sqrt 2 r^{3/2} Y_{1} + Y_{2} r^{3} \right) (1\pm \cos \psi) 
\ea
Since $Y_{1}>0$ and $ Y_{2}<0$, to get a stable solution we requires $n$ to be odd. This corresponds to    $z_1=-\frac{r^{3/2}}{\sqrt{2}}$ as in \cite{Baumann2}. 

The nonzero eigenvalues of case (b) are the same as in case (a) with an overall minus sign.
So we see that for this case to be stable, we need $n$ to be even corresponding to  $z_1=-\frac{r^{3/2}}{\sqrt{2}}$. This indicates that $F$ and $k^{\bar 1 1}_{w}$ are the same as in case (a)
so the potential is exactly the same as in case (a).

Finally, for case (c) to be stable we require that
\ba
(-1)^{m+n}\sqrt 2  r^{3/2} Y_{1} - r^{3} Y_{2} >0
\quad , \quad 
Y_{1} \sin^2 \theta_{1}  \left(  \sqrt 2 r^{3/2} Y_{1} - (-1)^{m+n} r^{3} Y_{2} \right) >0
\ea
To satisfy these conditions, we need $m+n$ to be even. This in turn implies that
 $z_1=-\frac{r^{3/2}}{\sqrt{2}}$  and the potential is the same as in case (a) and (b).



\begin{thebibliography}{99}


\bibitem{Klebanov:2000hb}
  I.~R.~Klebanov and M.~J.~Strassler,
  ``Supergravity and a confining gauge theory: Duality cascades and
  chiSB-resolution of naked singularities,''
  JHEP {\bf 0008}, 052 (2000)
  [arXiv:hep-th/0007191].

\bibitem{GKP}
  S.~B.~Giddings, S.~Kachru and J.~Polchinski,
  ``Hierarchies from fluxes in string compactifications,''
  Phys.\ Rev.\  D {\bf 66}, 106006 (2002)
  [arXiv:hep-th/0105097].

\bibitem{KKLT}
  S.~Kachru, R.~Kallosh, A.~Linde and S.~P.~Trivedi,
  ``De Sitter vacua in string theory,''
  Phys.\ Rev.\  D {\bf 68}, 046005 (2003)
  [arXiv:hep-th/0301240].

\bibitem{KKLMMT}
  S.~Kachru, R.~Kallosh, A.~Linde, J.~M.~Maldacena, L.~P.~McAllister and S.~P.~Trivedi,
  ``Towards inflation in string theory,''
  JCAP {\bf 0310}, 013 (2003)
  [arXiv:hep-th/0308055].

\bibitem{BKQ}
  C.~P.~Burgess, R.~Kallosh and F.~Quevedo,
  ``de Sitter string vacua from supersymmetric D-terms,''
  JHEP {\bf 0310}, 056 (2003)
  [arXiv:hep-th/0309187].

\bibitem{BCDF}
  C.~P.~Burgess, J.~M.~Cline, K.~Dasgupta and H.~Firouzjahi,
  ``Uplifting and inflation with D3 branes,''
  JHEP {\bf 0703}, 027 (2007)
  [arXiv:hep-th/0610320].

\bibitem{dvali-tye} G.~Dvali and S.-H.H.~Tye,
"Brane Inflation",
Phys. Lett. {\bf B450} (1999) 72, hep-ph/9812483.

\bibitem{collection}
C.~P.~Burgess, M.~Majumdar, D.~Nolte, F.~Quevedo, G.~Rajesh 
and R.~J.~Zhang, JHEP {\bf 07} (2001) 047, hep-th/0105204\, ; 
G.~R.~Dvali, Q.~Shafi and S.~Solganik,
``D-brane inflation,''
hep-th/0105203.    

\bibitem{HenryTye:2006uv}
  S.~H.~Henry Tye,
  ``Brane inflation: String theory viewed from the cosmos,''
  Lect.\ Notes Phys.\  {\bf 737}, 949 (2008)
  [arXiv:hep-th/0610221]; 
  J.~M.~Cline,
  ``String cosmology,'' arXiv:hep-th/0612129; 
  C.~P.~Burgess,
  ``Lectures on Cosmic Inflation and its Potential Stringy Realizations,''
  PoS {\bf P2GC}, 008 (2006)
  [Class.\ Quant.\ Grav.\  {\bf 24}, S795 (2007)]
  [arXiv:0708.2865 [hep-th]]; 
  L.~McAllister and E.~Silverstein,
  ``String Cosmology: A Review,''
  Gen.\ Rel.\ Grav.\  {\bf 40}, 565 (2008)
  [arXiv:0710.2951 [hep-th]].


\bibitem{Firouzjahi:2003zy}
  H.~Firouzjahi and S.~H.~H.~Tye,
  ``Closer towards inflation in string theory,''
  Phys.\ Lett.\  B {\bf 584}, 147 (2004)
  [arXiv:hep-th/0312020].

\bibitem{Shandera:2004zy}
  S.~E.~Shandera,
  ``Slow roll in brane inflation,''
  JCAP {\bf 0504}, 011 (2005)
  [arXiv:hep-th/0412077].

\bibitem{McAllister:2005mq}
  L.~McAllister,
  ``An inflaton mass problem in string inflation from threshold corrections  to
  volume stabilization,''
  JCAP {\bf 0602}, 010 (2006)
  [arXiv:hep-th/0502001].


\bibitem{Baumann1}
D. Baumann, A. Dymarsky, I.R. Klebanov, J. Maldacena, L.
McAllister and A. Murugan, ``On D3-brane potentials in
compactifications with fluxes and wrapped D-branes,'' JHEP {\bf
0611} 031 (2006) [hep-th/0607050].

\bibitem{Baumann2}
D. Baumann, A. Dymarsky, I.R. Klebanov and L. McAllister,
``Towards an explicit model of D-brane inflation,''
[arXiv:0706.0360 (hep-th)].


\bibitem{BBHL}
  K.~Becker, M.~Becker, M.~Haack and J.~Louis,
  ``Supersymmetry breaking and alpha'-corrections to flux induced
  potentials,''
  JHEP {\bf 0206}, 060 (2002)
  [arXiv:hep-th/0204254].


\bibitem{BB}
  V.~Balasubramanian and P.~Berglund,
  ``Stringy corrections to K\"ahler potentials, SUSY breaking, and the
  cosmological constant problem,''
  JHEP {\bf 0411}, 085 (2004)
  [arXiv:hep-th/0408054].

\bibitem{Cremades}
D. Cremades, M.-P. Garcia del Moral, F. Quevedo and K. Suruliz,
``Moduli stabilization and de Sitter string vacua from magnetized
D7 branes,'' [hep-th/0701154].

\bibitem{Misra:2007yu}
  A.~Misra and P.~Shukla,
  arXiv:0707.0105 [hep-th].
  A.~Misra and P.~Shukla,
  Nucl.\ Phys.\  B {\bf 800}, 384 (2008)
  [arXiv:0712.1260 [hep-th]].

\bibitem{Ouyang:2003df}
  P.~Ouyang,
  ``Holomorphic D7-branes and flavored N = 1 gauge theories,''
  Nucl.\ Phys.\  B {\bf 699}, 207 (2004)
  [arXiv:hep-th/0311084].

\bibitem{Kuperstein:2004hy}
  S.~Kuperstein,
  ``Meson spectroscopy from holomorphic probes on the warped deformed
  conifold,''
  JHEP {\bf 0503}, 014 (2005)
  [arXiv:hep-th/0411097].


\bibitem{Krause:2007jk}
  A.~Krause and E.~Pajer,
  ``Chasing Brane Inflation in String-Theory,''
  arXiv:0705.4682 [hep-th].

\bibitem{Pajer:2008uy}
  E.~Pajer,
  ``Inflation at the Tip,''
  JCAP {\bf 0804}, 031 (2008)
  [arXiv:0802.2916 [hep-th]].

\bibitem{Chen:2008ad}
  H.~Y.~Chen, J.~O.~Gong and G.~Shiu,
  ``Systematics of multi-field effects at the end of warped brane inflation,''
  arXiv:0807.1927 [hep-th].

\bibitem{Dasgupta:2002ew}
  K.~Dasgupta, C.~Herdeiro, S.~Hirano and R.~Kallosh,
  ``D3/D7 inflationary model and M-theory,''
  Phys.\ Rev.\  D {\bf 65}, 126002 (2002)
  [arXiv:hep-th/0203019];
  K.~Dasgupta, J.~P.~Hsu, R.~Kallosh, A.~Linde and M.~Zagermann,
  ``D3/D7 brane inflation and semilocal strings,''
  JHEP {\bf 0408}, 030 (2004)
  [arXiv:hep-th/0405247].

\bibitem{Haack}
  M.~Haack, R.~Kallosh, A.~Krause, A.~Linde, D.~Lust and M.~Zagermann,
  ``Update of D3/D7-Brane Inflation on $K3 \times T^2/Z_2$,''
  arXiv:0804.3961 [hep-th].

\bibitem{Chen:2004gc}
  X.~Chen,
  ``Multi-throat brane inflation,''
  Phys.\ Rev.\  D {\bf 71}, 063506 (2005)
  [arXiv:hep-th/0408084]; 
  X.~Chen,
  ``Inflation from warped space,''
  JHEP {\bf 0508}, 045 (2005)
  [arXiv:hep-th/0501184]; 
  S.~Thomas and J.~Ward,
  ``IR Inflation from Multiple Branes,''
  Phys.\ Rev.\  D {\bf 76}, 023509 (2007)
  [arXiv:hep-th/0702229];
  J.~E.~Lidsey and I.~Huston,
  JCAP {\bf 0707}, 002 (2007)
  [arXiv:0705.0240 [hep-th]];
  R.~Bean, X.~Chen, H.~V.~Peiris and J.~Xu,
  ``Comparing Infrared Dirac-Born-Infeld Brane Inflation to Observations,''
  Phys.\ Rev.\  D {\bf 77}, 023527 (2008)
  [arXiv:0710.1812 [hep-th]].
  B.~Underwood,
  ``Brane Inflation is Attractive,''
  arXiv:0802.2117 [hep-th].



\end{thebibliography}
\end{document}